\documentclass{astlb} 

\usepackage[hyperfootnotes=false]{hyperref}

\usepackage{color}
\usepackage{epsf}

\usepackage[utf8]{inputenc}
\usepackage[T2A]{fontenc}

\usepackage{amsmath}
\usepackage{amsfonts}
\usepackage{amssymb}

\usepackage{epsf}
\usepackage{graphicx}
\usepackage{gensymb}
\usepackage{float}
\usepackage{changes}

\def\beq#1{\begin{equation}\label{#1}}
\def\eeq{\end{equation}}
\def\beqa#1{\begin{eqnarray}\label{#1}}
\def\eeqa{\end{eqnarray}}

\def\comment#1{\relax}

\def\apgt{\ {\raise-.5ex\hbox{$\buildrel>\over\sim$}}\ }
\def\aplt{\ {\raise-.5ex\hbox{$\buildrel<\over\sim$}}\ }

\begin{document}

\journalinfo{2020}{46}{11}{1}[18]

\title{Transient Double-beam Spectrograph for the 2.5-m Telescope of the Caucasus Mountain Observatory of SAI MSU}

\author{
  S.~A.~Potanin \email{potanin@sai.msu.ru}
  \address{1,2},
  A.~A.~Belinski\address{2},
  A.~V.~Dodin\address{2},
  S.~G.~Zheltoukhov\address{1,2},
  V.~Yu.~Lander\address{1},
  K.~A.~Postnov\address{2,1},
  A.~D.~Savvin\address{2},
 A.~M.~Tatarnikov\address{2},
  A.~M.~Cherepashchuk\address{2},
  D.~V.~Cheryasov\address{2},
  I.~V.~Chilingarian\address{3,2},
  N.~I.~Shatsky\address{2}  \email{potanin@sai.msu.ru, kolja@sai.msu.ru}.
\addresstext{1}{Faculty of Physics, Lomonosov Moscow State University, Leninskie Gory 1-2,  119991 Moscow, Russia}
 \addresstext{2}{Sternberg Astronomical Institute, Universitetskij pr. 13, 119234 Moscow, Russia}
\addresstext{3}{Smithsonian Astrophysical Observatory, 60 Garden St MS09, Cambridge, MA, 02138, USA}
}

\shortauthor{S.A. Potanin et al.}

\shorttitle{TDS spectrograph for CMO SAI MSU}

\submitted{27.10.2020}

\begin{abstract}
  
The Transient Double-beam Spectrograph (TDS) is designed for optical low-resolution observations of non-stationary and extragalactic sources with the 2.5-m telescope of Caucasus Mountain Observatory of the Sternberg Astronomical Institute. It operates simultaneously in a short-wavelength (360--577~nm, reciprocal dispersion 1.21~\AA/pixel, resolving power R=1300 
with a 1\arcsec-wide slit) and long-wavelength (567--746~nm, 0.87 \AA/pixel, R=2500) channels. The light is split by a dichroic mirror with a 50\% transmission at 574~nm. In the ``blue'' channel, the automatic replacement of the grating by a grism with a double resolving power is possible.  
Two CCD-cameras use E2V 42-10 detectors cooled down to $-70^{\degree}$C  with a readout noise of 3~$e-$ at a readout rate of 50 kHz. The spectrograph is equipped with a back slit viewer camera and a calibration unit allowing to record a comparison  spectrum from a hollow cathode lamp for wavelength calibration or from an LED source with a continuous spectrum (the ``flat field'') to take into account the vignetting and uneven slit illumination. 
The throughput of the entire optical path without slit loss is 20\% at the zenith in the “blue” channel and 35\% in the “red” channel. Excluding the atmosphere and the telescope, the efficiency of the TDS itself reaches a maximum of 47\% and 65\% respectively. The spectrograph is permanently mounted in the Cassegrain focus of the 2.5-m telescope of CMO SAI MSU sharing the port with a wide-field photometric CCD-camera. The spectrograph is fed by the light from a folding mirror introduced into the optical path.
Since November 2019, TDS has been used for regular observations of non-stationary stars and extragalactic sources up to $\sim\!20^{\rm m}$ in a 2-h exposure with a signal-to-noise ratio $>5$~pix$^{-1}$.


\end{abstract}


\section*{Introduction}
\label{s:intro}
Modern astrophysical research of transient sources, such as search and classification of supernovae, optical afterglow of gamma-ray bursts, the study of non-stationary phenomena in close binary systems with relativistic components, and the physics of pre-Main Sequence objects, require prompt acquisition and classification of optical spectra with intermediate-size telescopes in the follow-up and monitoring modes. The ability to rapidly respond to a new observing request and high efficiency of observations are the most critical factors; therefore, the effectiveness of such projects directly depends on the total throughput of the optical system and the possibility of scheduling observations at any moment. Despite the large number of intermediate-size and large telescopes that were put into operations in the recent decades, the quick characterization of astrophysical transients remains a scientific problem of current interest \citep{2019MNRAS.489.1463O, 2015ExA....39..119C, 2016MNRAS.462.3528C}.

The telescope of the Caucasus Mountain Observatory (CMO) of SAI MSU with a 2.5-m primary mirror is a versatile instrument for scientific and educational purposes \citep{2020arXiv201010850S}. The observational programs are chosen from a wide range of research topics on a competitive basis by the time allocation committee of SAI in order to achieve the maximum final efficiency of the facility taking into account its capabilities and site characteristics. Weather and seeing conditions at CMO SAI \citep{2014PASP..126..482K} and the possibility of their rapid assessment allow for flexible planning of observations with a number of the facility instruments based on changing observing conditions, and thus the most efficient use of the telescope time. The telescope was installed in 2014 and is currently (November 2020) preparing for final commissioning tests after replacing the hardware and software parts of the telescope control system.

To support the characterization of new sources and spectral monitoring of non-stationary objects conducted at SAI MSU (see, for example, \citealt{2016A&A...588A..90L, 2017MNRAS.467.3500V, 2018ARep...62..747C,  2020arXiv200807934B}), an effective instrument for spectral classification and measurement of Doppler shifts and intensities of emission and absorption lines was required. The high throughput of such spectrographs for 1.5--2m class telescopes is crucial for their high productivity \citep[see, e.g. the overview of observing programs at the 1.5-m telescope in][and references therein]{2020ASPC..522..655M}.

The relatively large CMO SAI telescope aperture and sub-arcsec image quality \citep{2017ARep...61..715P} make it possible to use a narrow and long slit for long-slit measurements of extended sources, for example, galaxies, with the ability to accurately measure the sky background spectrum. Therefore, the instrument aims at having a higher resolving power $R\!\sim\!2000--3000$ and a longer slit than, for example, the FLOYDS spectrographs of the 2-m telescopes of the Las Cumbres observatory ($R<1000$, $L_{slit} \! = \! 30$\arcsec; \citealt{2013PASP..125.1031B}), specially designed for observing transient objects. In terms of the slit parameters, spectral resolution, and the diversity of typical scientific projects, the proposed spectrograph is more reminiscent of the Intermediate Dispersion Spectrograph\footnote{\url{http://www.ing.iac.es/Astronomy/instruments/ids/}} at the 2.5-m INT telescope in the Canary Islands (see, for example, \citealt{2019A&A...627A.124P}), although the stability requirements are poorly combined with such a large set of dispersors and interchangeable detectors. Therefore, from the very beginning, the Transient Double Spectrograph was designed as an instrument with a minimal acceptable number of configurations.

\section{Criteria for Choosing the Characteristics of the TDS}
\label{sec:criteria}

Although the 2.5-m SAI telescope is a  versatile instrument, its moderate aperture size requires specific design features for the spectrograph to achieve the maximal performance for its main purpose, the spectroscopy of 20th-magnitude transient sources and extragalactic objects. Therefore, the primary criterion for optimizing the design is a high throughput combined with a certain minimally acceptable value of the spectral resolving power and a broad enough wavelength coverage required for successful object classification. The modern lines of optical glasses available on the market and the widespread introduction of holographic volume phase diffraction gratings (VPH, \citealt{2000PASP..112..809B}) into astronomical practice allow us to to solve both these problems with the minimal light loss in the optical system.

The spectral range required for typical observing programs spans from the Balmer break up to the end of the visible. We also need a sufficiently long (3\arcmin) slit to assure accurate sky background subtraction and the possibility of observing extended sources. Both requirements are in good agreement with the format of available highly efficient spectral CCD cameras capable of long-term operation in the near-cryogenic mode without significant maintenance costs for the system operating in the `constant alert' mode. The spectral resolving power of 1500--2500 required for spectral classification and measurements of energy fluxes in spectral lines, turns out to be achievable using a double-beam design of the instrument with a channel separation by a dichroic beam splitter at the wavelength between 550 and 580~nm. This approach eliminates the need of using order blocking filters and allows the use of efficient antireflection (AR) coatings for optics and CCDs, as well as achromatization of optics with a small number of elements. This solution is similar to the layout of the late generation SDSS BOSS spectrographs\footnote{\url{https://www.sdss.org/instruments/boss\_spectrograph/}} \citep{2013AJ....146...32S}, but scaled by the detector size taking into account with a long slit and a narrower spectral range in the red domain. The latter feature made it possible to use a common collimator for the two channels with minimal compromises in the image quality. This approach to the optical design made the device more compact and containing fewer optical elements than the similar spectrograph FRODOSpec of the Liverpool 2-m telescope with two dioptric collimators \citep{2004AN....325..215M}.

The final multiplexing of the spectrograph is defined by a nearly complete coverage of the area of the two CCDs with a format of 2048$\times$512~pix each by a spectrum covering 350$<\lambda<$750~nm. The wavelength range sampled at $R\!\sim\!2000$ with FWHM$=$3~pix for a 1\arcsec-wide slit yields the slit length of about 3\arcmin. These requirements are consistent with the median seeing at the site of $1.0\arcsec$ and also the image quality provided by the telescope optics \citep{2017ARep...61..715P}. Optimization of the scheme on a wide range of modern optical glasses made it possible to obtain the required image quality in the entire spectral and spatial range with a small number of elements and high spectral transmission, unattainable using commercial off-the-shelf lenses.

The required spectral resolution is achieved by using VPH gratings optimized for the Bragg angle for the middle of their working regions in the channels. The main dispersors are used without compensating prisms, with a bend in the optical axis of the instrument. Besides the increased throughput, this lowers the weight and dimensions of the Cassegrain instrument, provided that the direct image mode, which is common for focal reducer spectrographs \citep{1984Msngr..38....9B, 2005AstL...31..194A}, is not required because a separate CCD photometer is available in the same focus. Such a solution was proposed at the dawn of VPH-astrospectroscopy in the ATLAS project \citep {2000ASPC..195..110T}.

Identification, quasi-simultaneous photometry, and alignment of faint source on the slit are performed using a wide-field CCD photometer (a $4K\!\times\!4K$ pixels camera having  a 10\arcmin$\times$10\arcmin\ field of view; \citealt{2020arXiv201010850S}). The camera operates in the same port of the telescope in the direct viewing mode, while the light is fed into the TDS by a folding mirror introduced in the optical path. Such a ``tandem'' makes it possible to avoid achromatization of the spectrograph optical system in the imaging mode which greatly simplifies the optical design and gives a significant gain in the total throughput.

The list of requirements is completed by the athermal design of camera lens cells, the presence of an optical calibration system (mirror-lens system of image transfer of the sources of arc and continuum spectra) and also providing an extra mode with a double resolving power in the long-wave (``green'') part of the blue channel. These requirements are common for similar instruments discussed above.

In the following section, we discuss the details about the optical and mechanical design of the spectrograph.

\section{Optical Design}

\begin{figure*}
\includegraphics[width=\hsize]{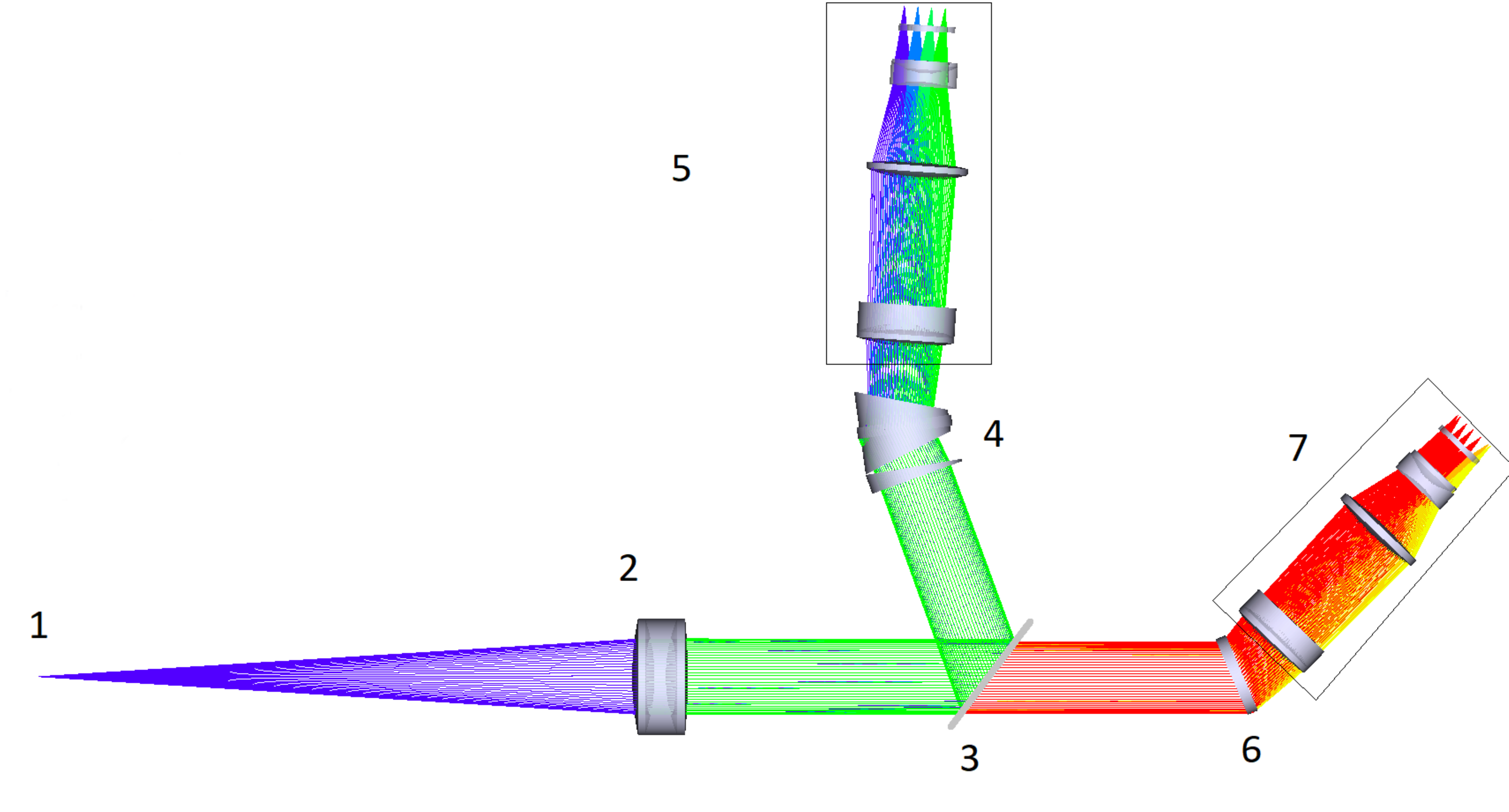}
\caption{Optical layout of the TDS spectrograph. 1 -- slit in the focal plane of the telescope, 2 -- common collimator, 3 -- flat dichroic mirror, 4 -- replaceable blue channel dispersor ($G$ grism and an additional wedge are shown), 5,7 -- blue and red channel camera lenses, 6 -- red channel dispersor.}
\label{fig:TDS_opt}
\end{figure*}

Fig.~\ref{fig:TDS_opt} shows the optical layout of the instrument. The light enters the spectrograph's slit by a reflection from an aluminized folding mirror introduced into the Cassegrain optical axis. After the slit, the light enters the collimator triplet. Between the collimator and the pupil of the system, there is a beam splitter that reflects light with a wavelength shorter than $0.57\mu m$ into the blue channel of the spectrograph and passes longer waves into the red channel. A non-standard operating angle of incidence of light, $35^o$, was chosen for the optical and mechanical design optimization. The two cameras follow immediately behind the dispersors installed in the pupils.

\begin{figure*}
\includegraphics[width=0.49\hsize]{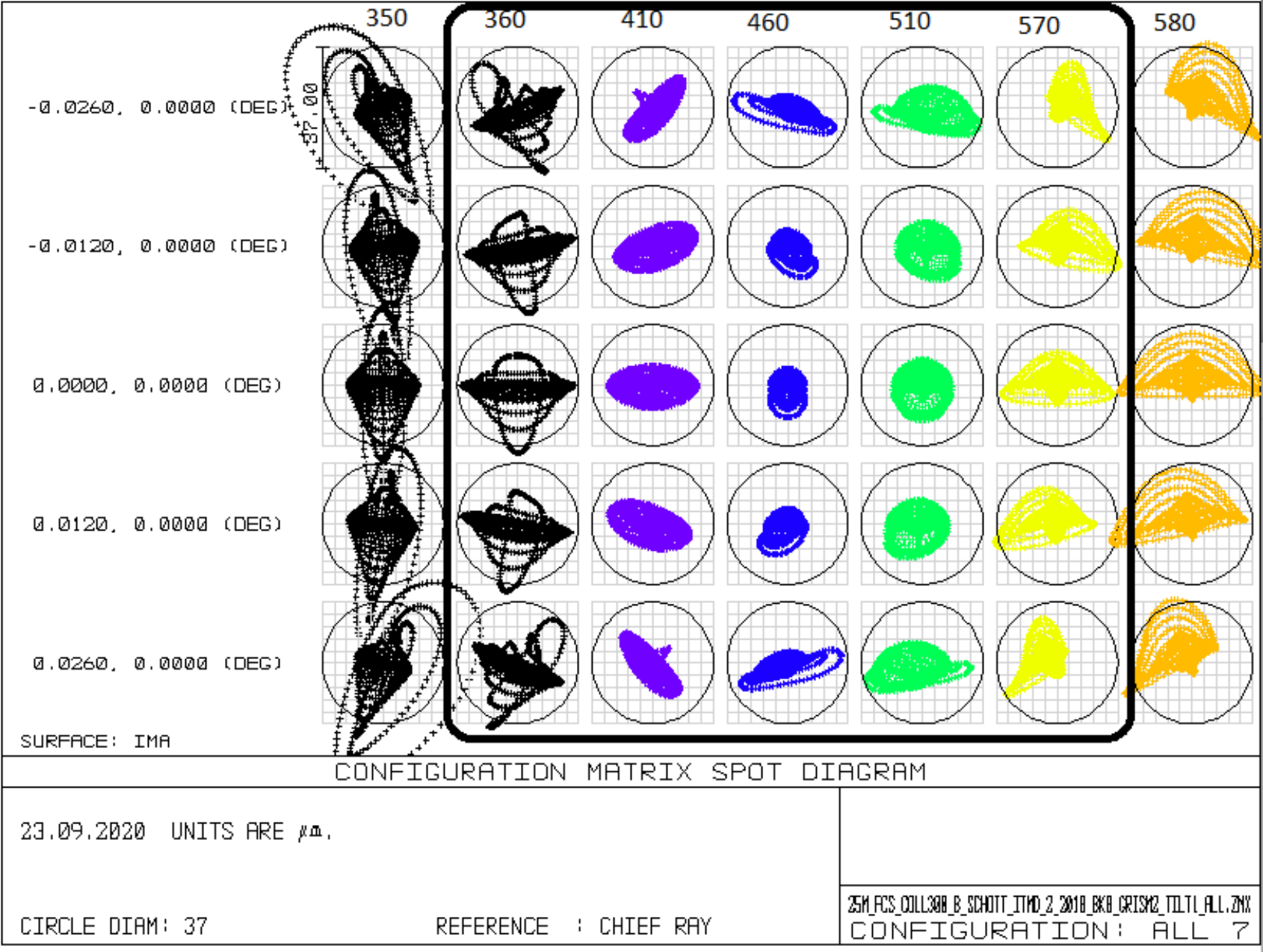} \includegraphics[width=0.49\hsize]{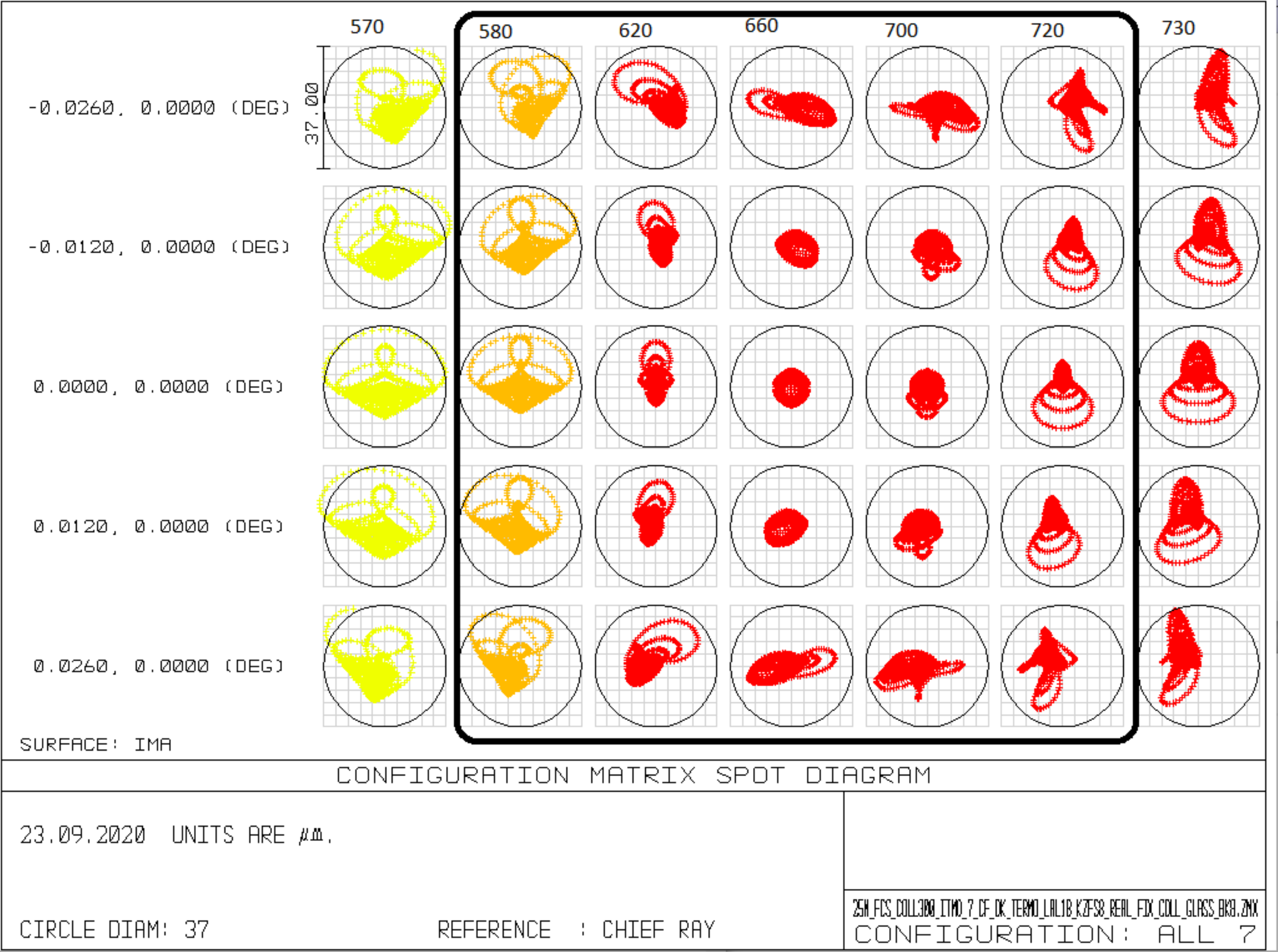}
\caption{Ray diagrams in the blue channel (left; main $B$ grating) and in the red channel $R$ (right) for different wavelengths (horizontal) and through slit sections (vertical). The circle corresponds to the image diameter of 1~arcsec in the focal plane of the telescope. The working area of the wavelength ranges are marked with a black rectangle. The wavelengths in nm are indicated above the corresponding image columns.}
\label{fig:TDS_Image}
\end{figure*}

The optical scheme was calculated in the {\sc zemax} software in the multiconfiguration mode, separately for the red and blue channels. With a minimum number of groups of elements (one in the collimator and two doublets plus a single lens in each of the cameras), the energy concentration in the images of 80\%\ was achieved in a circle with a diameter of 24 microns (approximately evenly along the slit and along the dispersion). The calculated images are shown in Fig.~\ref{fig:TDS_Image}. The tolerances for the manufacture and alignment of the optics turned out to be moderately loose (0.1~mm in the shift of elements).

The main parameters and elements of the spectrograph are summarized in Table~\ref{tab:param}. The slits are laser cut from stainless steel 200~microns thick on a robotic machine for cutting focal masks of the Binospec instrument and are kindly provided by the D.~Fabricant (Smithsonian Astrophysical Observatory, USA; \citealt{2019PASP..131g5004F}), their surface is not blackened. In the wheel of focal apertures, a working slit with a width of 0.1~mm (1\arcsec; scale in the focal plane of the telescope is 10\arcsec/mm), a field diaphragm with a diameter of 18~mm for object acquisition (also used for slitless spectroscopy), as well as a wide spectrophotometric slit are installed.

\begin{table}
\caption[Parameters of elements]{\label{tab:param}
Basic parameters of elements and output spectral characteristics
}
\begin{tabular}{r|l}
Parameter & Value\\
\hline
{\bf Slits} \\
Width & 1\arcsec and 10\arcsec \\
Height & 3\arcmin\  (18~mm)\\
{\bf Collimator} \\
Focal length & 315 mm \\
Lens diameter & 64 mm \\
{\bf Dichroic splitter} \\
Incidence angle & $35^{\degree}$ \\
Dimensions & 65x90 mm \\
Reflectance cut-off & 350 nm \\
Average reflectance & 98.2\% \\
Cut-On wavelength & 574 nm \\
Transmission cut-off & 750 nm \\
Average transmission & 95.3\% \\
{\bf VPH dispersors} \\
Pupil size & 39 mm \\
Light diameter & 43 mm \\
Ruling density (R,B,G) & 1200,900,1800~lpmm \\
Central wavelength & 650, 460, 505 nm \\ 
{\bf Camera lenses} \\
Focal length & 117 mm \\
Field of view angle, $2\Omega$ & $13.9^{\degree}$ \\
{\bf Detectors} \\
CCD Model & E2V 42-10 \\
Pixel size & 13.5x13.5 $\mu$ \\
Quantum efficiency & 90\% \\
Readout rate & 50 kHz \\
Temperature setpoint & $-70^{\degree}$C \\
Readout noise (R, B) & 3.8, 3.1 e- \\
{\bf Blue channel} \\
Reciprocal dispersion ($B$, $G$) & 1.21, 0.55\AA/pix \\
Wavelength range in $B$ & 3600--5770\AA \\
Wavelength range in $G$ & 4300--5434\AA \\
Resolving power R ($B$, $G$) & 1300, 2600 \\
Max. efficiency ($B$, $G$) & 47\%, 50\% \\
{\bf Red channel} \\
Reciprocal dispersion & 0.87\AA/pix \\
Wavelength range in R & 5673--7460\AA \\
Resolving power R & 2500 \\
Max. efficiency ($R$) & 65\% \\
\hline
\end{tabular}
\end{table}

The key component of a two-channel spectrograph is a dichroic beam splitter. It is installed ahead of the pupil of the system, since the pupil is occupied by dispersive elements, which determine the spectral characteristics of the channels. The main requirements for the splitter are to provide high throughput in the long wavelength range and high reflection in the short one. Another important features are the steepness of the spectral response around the cut-off wavelength and as low as possible amplitude of spectral response oscillations (``waves'') about the the average response level unavoidable in such systems. This determines the spectral profile of the total throughput of the instrument and the possibility of correcting final calibrated spectra for its possible time variability to obtain accurate spectrophotometric characteristics of astrophysical sources.

The initial reference calculation of the dichroic beam splitter was kindly performed by Asahi Spectra\footnote{\url{http://www.asahi-spectra.com}, the leading manufacturer of astronomical interference and glass filters} and showed the theoretical possibility of providing average losses within 3\% and the amplitude of the ``waves'' well under 5--10\%. However, for economic reasons, we could not order this mirror and a searched for a domestic manufacturer.

\begin{figure}
\includegraphics[width=0.49\textwidth]{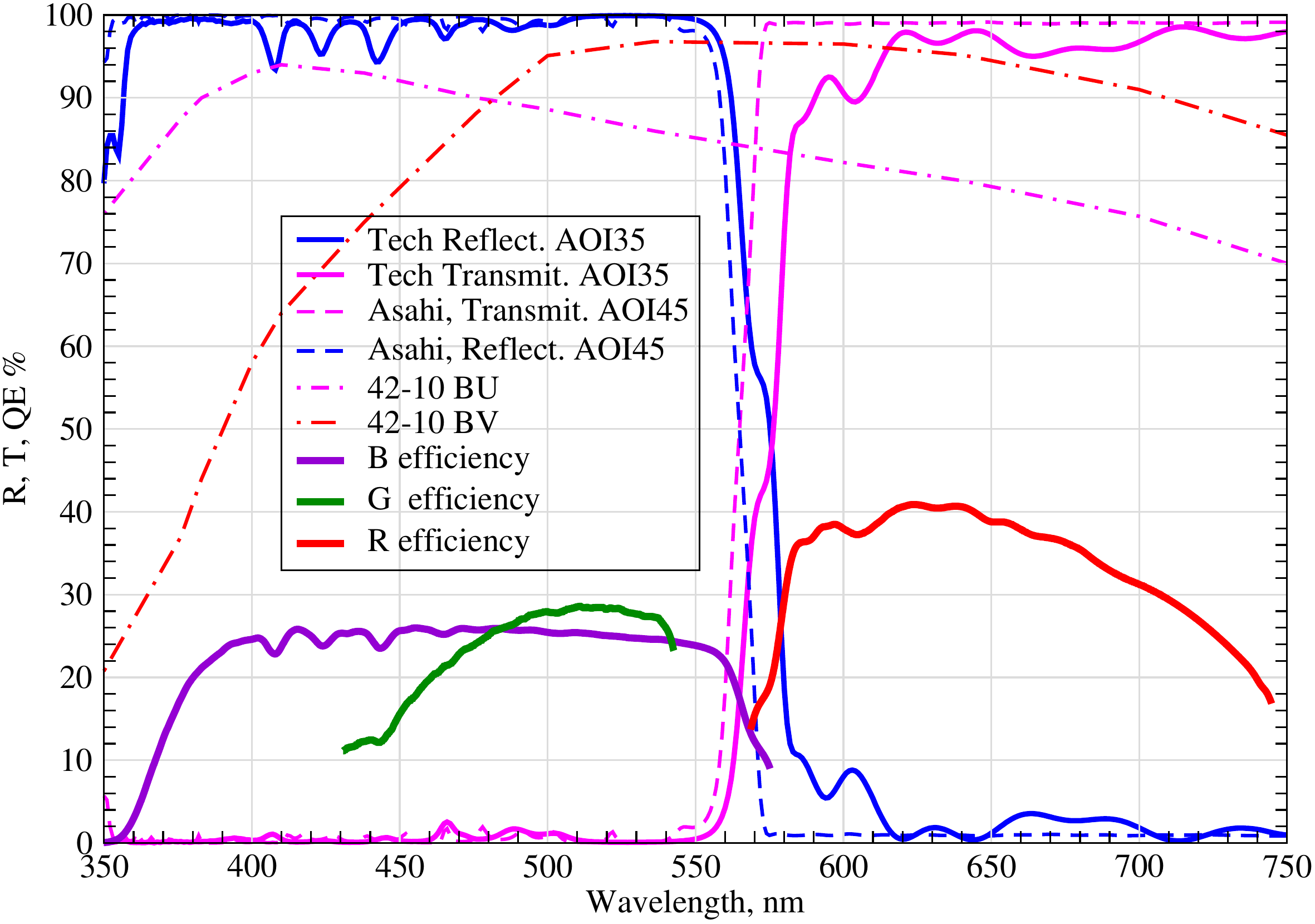}
\caption{The surface-averaged measured transmission and reflection curves of the working beam splitter (thin solid lines, incidence angle $ 35^o$) and reference curves of the dichroic divider calculated for the TDS in Asahi Spectra (dashed lines, for the incidence angle $45^o$); given half-sum for S- and P-polarizations. For clarity, the dashed-dotted line shows the curves of the quantum efficiency of the receivers, and the thick lines show examples of the final output spectral response curve of the entire system with the offset beyond the atmosphere.}
\label{fig:dichro}
\end{figure}

At the end, an acceptable design, and then experimentally confirmed characteristics of a dichroic coating were obtained at the Ryazan production site of NPP ``Alexander, LLC''\footnote{\url{https://macrooptica.ru}; the calculation was carried out by Technion, LLC}, which manufactured the beam splitter of the instrument. Its throughput curves are shown in Fig.~\ref{fig:dichro}, displayed in comparison to the theoretical performance of a coating from Asahi Spectra.

Diffraction gratings of the spectrograph with a diameter of 2 inches were manufactured by Wasatch Photonics\footnote{\url{https://wasatchphotonics.com/}} and they all operate in the first diffraction order; the main channel gratings have AR coatings optimized for their operating range. 

A red (`$R$') VPH grating on the B270 glass substrates is located in the pupil of the red channel deflecting the central wavelength by  $46^{\degree}$. The pupil of the blue channel hosts a block of two interchangeable VPH gratings on fused silica substrates, made specifically for our project. The main grating (`$B$') deflects the beam by $24^{\degree}$ and is used without additional prisms. The auxiliary (`$G$') grating with a double spectral resolving power and a central wavelength of 505~nm is used with the prisms made of the F1 flint glass (Lytkarino factory) to align the deflection angle with the $B$ grating. To eliminate the emerging aberrations and focus differences in the blue and green channels, the front prism is supplemented with a wedge made of KU-1 silica with a slightly convex front surface ($R\!=\!54$~m).

Both channels use similar camera lenses that are optimized for their wavelength ranges. The sorts of glass used in them\footnote{Schott, Ohara and LZOS glasses are used in the TDS optics, glued with OK-72FT5 low-temperature glue, transparent down to 350~nm}, and the AR coatings are different for the lenses in the red and blue channels. Five-element lenses have a back focal distance of 30~mm, which allows us to use a CCD camera of virtually any design, including an optical window and a shutter. We avoided the use of fluorite in the optics of the spectrograph thus eliminating the need for channel focusing and relaxing the temperature stresses; the images remain sharp in the temperature range from $-15^{\degree}$ to $+25^{\degree}$C.

The TDS spectrograph uses Andor Newton DU940P cameras\footnote{\url{https://andor.oxinst.com/products/newton-ccd-and-emccd-cameras/newton-940}} with central shutters and plane-parallel fused silica optical windows. The first grade back-illuminated arrays of the BU-specification for the blue channel (as in the FLOYDS instruments) and BV-spec for red channel operate with the 1.0~$e-$/ADU signal digitization at a constant operating temperature provided by thermoelectric air cooling all year round. The contribution of the dark current to the data noise is small and amounts to less than 0.005~$e-$~s$^{-1}$.

\section{Mechanical Design}
\label{sec:mech}

\begin{figure}
\includegraphics[width=0.47\textwidth]{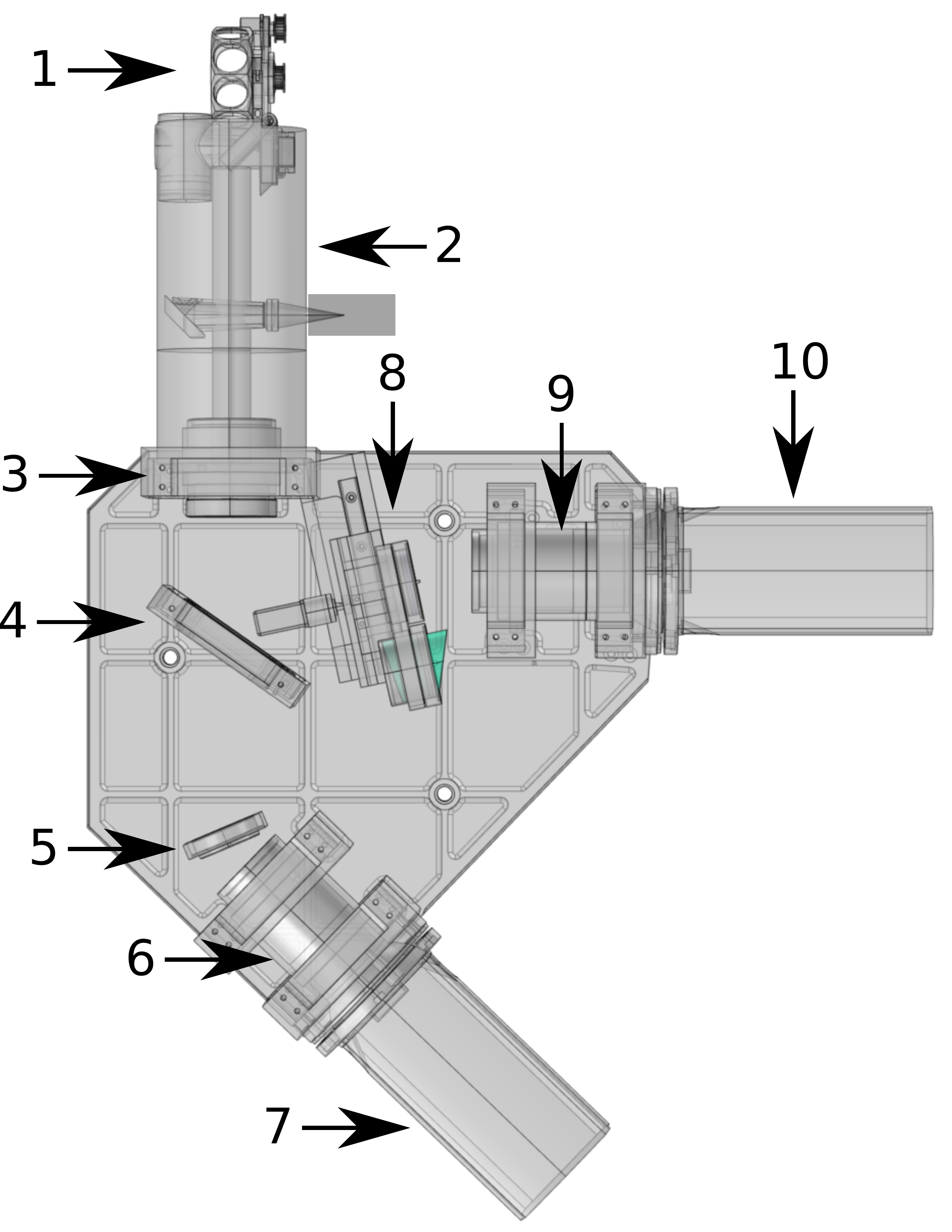}

\caption{Model of the mechanics of the spectrograph. 1 -- wheel of interchangeable slits, 2 -- tube with a image transfer system and a viewer camera, 3 -- collimator, 4 -- dichroic beam splitter, 5 -- red channel grating, 6,9 -- lenses of red and blue channels, 7,10 -- CCD cameras of red and blue channels, 8 -- block of interchangeable blue channel dispersors.}
\label{fig:mech}
\end{figure}

The spectrograph was designed and assembled at M.V.~Lomonosov Moscow State University, where some of the parts were also manufactured, the remaining components were ordered from outside suppliers. The main body of the device is made of deformable aluminum alloys (AMG-6 and D16) to prevent the leash during processing and consists of a 3~cm thick base plate with stiffening ribs, rectangular rigid stands for frames of optical components and flanges of CCD cameras, and the body cover, similar to the base, 2~cm thick. This design gives the necessary rigidity to the device and is complemented with side light-shielding covers. The spectrograph is fastened with screw pins through three axial holes of the additional housing legs to an aluminum angle support truss attached to the Cassegrain filters and shutter unit (FSU) of the wide-field telescope camera. The mechanical construction of the spectrograph is shown in Fig.~\ref{fig:mech}.

The block of interchangeable slits and a slit viewer is mounted in a tube fixed on the collimator stand. The 4-seat slit wheel is protected by a cover (not shown) and rotates in precision bearings so that it moves in the direction of the slit for stable and repeatable wavelength calibration. Immediately behind the slits, a folding flat mirror transfers the image of the input apertures with de-magnification to a CMOS slit-viewing  camera\footnote{Model BFLY-PGE-23S6M-C, \url{https://www.flir.eu}} to align objects on the slit. The wheel and mirror are driven by micro-servo motors.

The collimator and both camera lenses are manually focused when setting up the device and are clamped with locknuts. The lenses are mounted in sleeves with a slip fit and do not rotate during focusing. The red channel grating is tilted for maximum efficiency in the first order. The carriage of interchangeable blue channel dispersors driven by a stepper motor and a precision Omron D5A zero-point sensor also has beam angle adjustments. All dispersors are fine-tuned in the direction of the grooves by rotating around the optical axis.

The CCD cameras in both channels are mounted on adjusting flanges having a ball-shaped surface with a center of curvature on the surface of the array. This allows a slight rotation or tilt of the detector with respect to the optical axis without disturbing focus in the central part.

The light transfer and calibration system is mounted separately from the spectrograph inside the FSU. It is a linear guide with a backlash-free ball screw which carries a platform with diagonal mirrors\footnote{38$\times$54 mm Elliptical Mirror Protected Aluminum EO \#30-258, \url{https: //www.edmundoptics.com}}. One mirror, when set on axis, deflects the telescope light into the TDS. The other mirror is mounted in a pair with an achromatic doublet (EO \#49-286, F=200~mm). It telecentrically transfers the image of the output port of the integrating sphere (IS) containing calibration sources onto the slit. The carriage is moved by a stepper motor with an encoder and can also occupy the position (3) to let the light into the 4K-camera and the position (4) reserved for the mirror feeding a future high-resolution fiber-fed spectrograph.

The calibration sources include a hollow cathode gas-discharge lamp for wavelength calibration and a composite LED source with a continuum spectrum (``flat field''). Light from a hollow cathode lamp (HCL) is fed into the IS via an optical fiber, and the LED source is mounted directly on the IS.

\begin{figure*}
\includegraphics[width=\hsize]{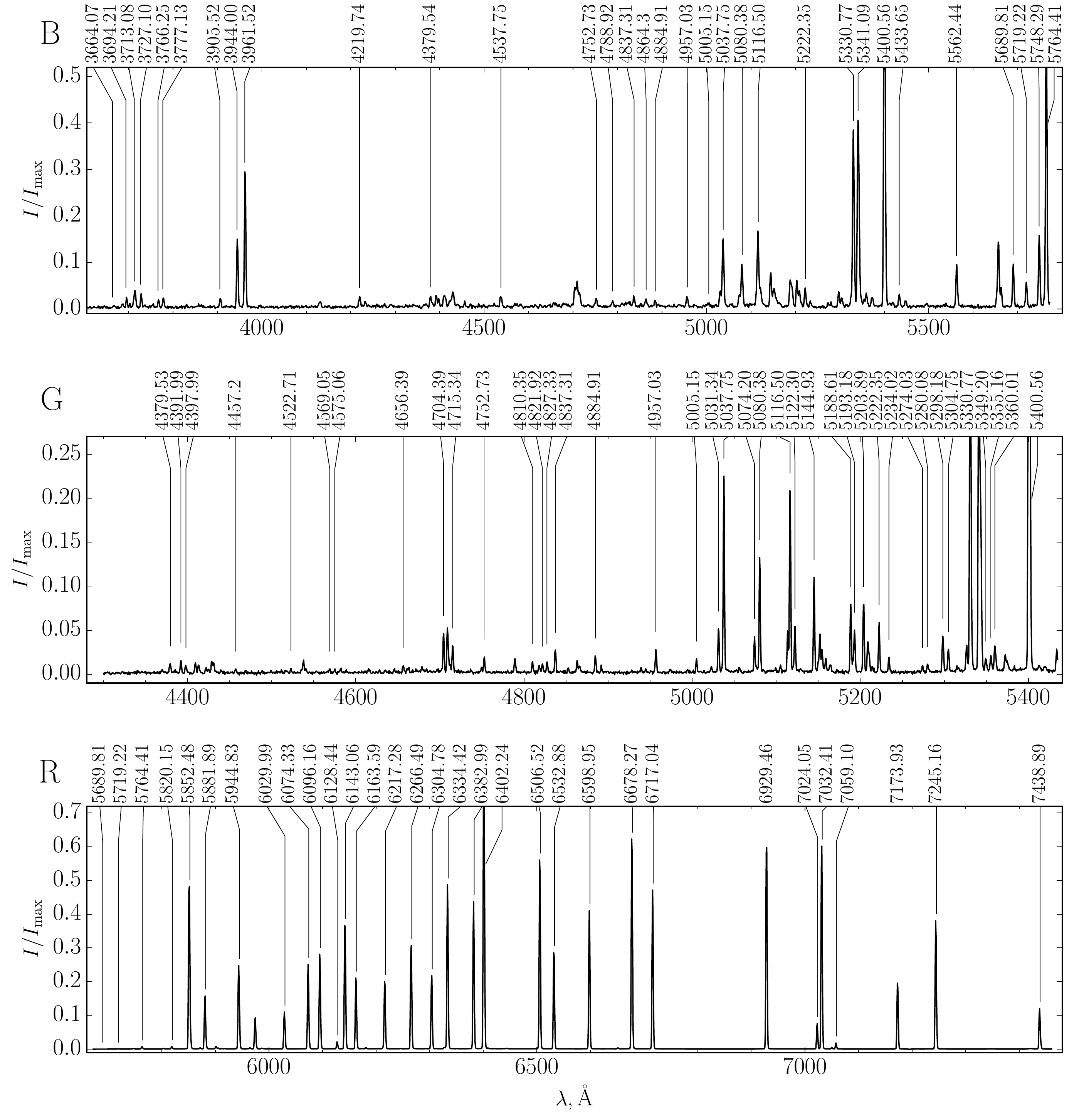}

\caption{Spectrum of the Ne+Al+Si calibration lamp with wavelengths of the lines used for the wavelength calibration}
\label{fig:lamp_bond2}
\end{figure*}

Until September 2020, a HCL containing Ne+Kr+Pb+Na was used, which was then replaced by the Ne+Al+Si lamp. Both lamps were made and kindly provided by Yu.N.~Bondarenko (Odessa observatory) and show a number of single bright lines in the $R$ channel and many partially blended weak lines in the $B$ channel (Fig.~\ref{fig:lamp_bond2}). This slightly complicates the calibration process, because in the $B$ channel it requires a much longer exposure (about 15~min) to obtain a lamp spectrum with a signal-to-noise ratio high enough to reconstruct all terms of the two-dimensional wavelength solution (the red channel exposure is 3--5 min).

The continuum spectrum source includes a broadband super-bright LED with a ``sun-like'' spectrum\footnote{Model STW9C2PB-S Q54CY3, see \url{http://www.seoulsemicon.com}} as well as additional shorter wavelength 402, 380 and 365~nm LEDs whose flux is adjusted by variable resistors. The current and temperature of the substrate are stabilized by the on-board electronics and retain the long-term relative spectral brightness of the diodes within 0.02\% after a 1-minute warm-up. The use of xenon lamps was abandoned due to the presence of bright sharp bands in the spectrum.

The control and read-out of the CCD cameras is performed in the Linux operating system using the SDK library version 2.102.30024. The acquisition software is implemented in C/C++ using an API for Andor devices developed at Sternberg Astronomical Institute. It saves the data together with the metadata of observation circumstances in FITS files. The slit wheel, the viewer mirror and blue dispersor carriage motors are driven by the Arduino controllers, which also monitor the digital temperature sensors of the optics cells. The control computer of the TDS spectrograph is installed in the server room of the 2.5-m telescope tower and communicates with CCD cameras and controllers via a 4-channel fiber-optic USB extension cable. The software to control the spectrograph and communicate with the telescope control system is implemented in {\sc Python}.


\section{Flexures of the Instrument and Stability of the Wavelength Solution}
\label{sec:deform}
Like many Cassegrain instruments, the TDS is not thermally stabilized and during observations it changes the orientation in the course of pointing and tracking of astrophysical sources, as well as when the position angle of the slit is adjusted by the port rotator. The deformations from gravity and temperature, when conditions change, affect its wavelength calibration. If one does not take into account the corresponding offsets, it can lead to systematic errors in the measured radial velocities of astrophysical sources. We use a calibration lamp with arc lines and night sky emission lines to assess the amplitude of the effect.

The TDS uses transmission gratings and a grism, therefore the dispersion relation in the first approximation is linear. For calibration, a polynomial of the 5th order is used to account for optical distortion and compensate the dispersion by prisms; a further increase in the degree does not lead to a significant improvement in the quality of the wavelength solution. The nonlinear terms of the dispersion relation have a relative full amplitude of 2\% in $B$, 1.4\% in $R$ and 0.3\% in $G$.

In the spatial direction, a distortion of up-to 0.3\% is observed as well as bending and tilt of the slit images of up-to 5 pixels (in the red channel) with respect to the middle line of the spectrum image. To correct spatial distortions the polynomials of the 3rd degree are used.

\begin{figure}
\includegraphics[width=0.49\textwidth]{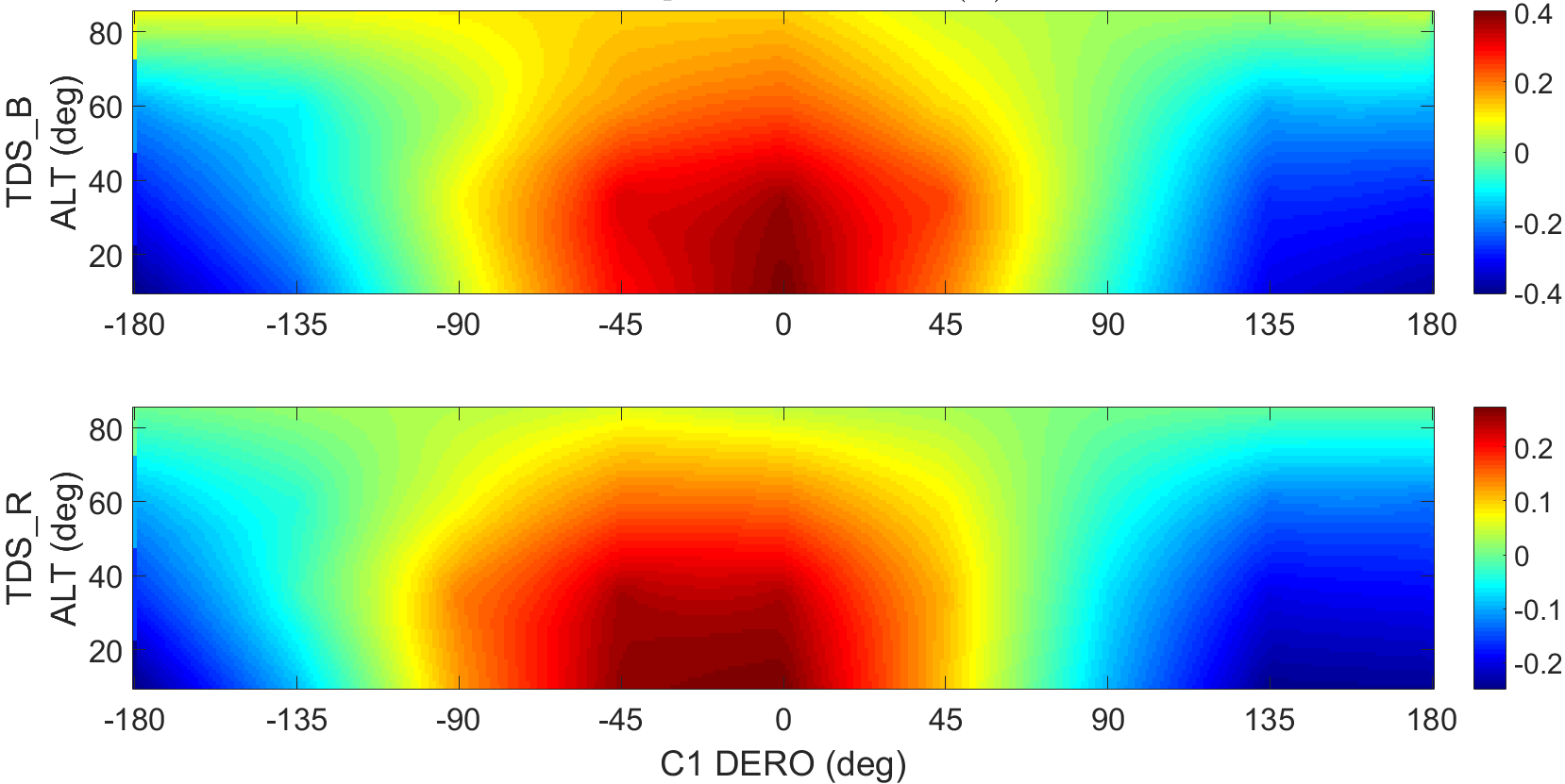}
\includegraphics[width=0.49\textwidth]{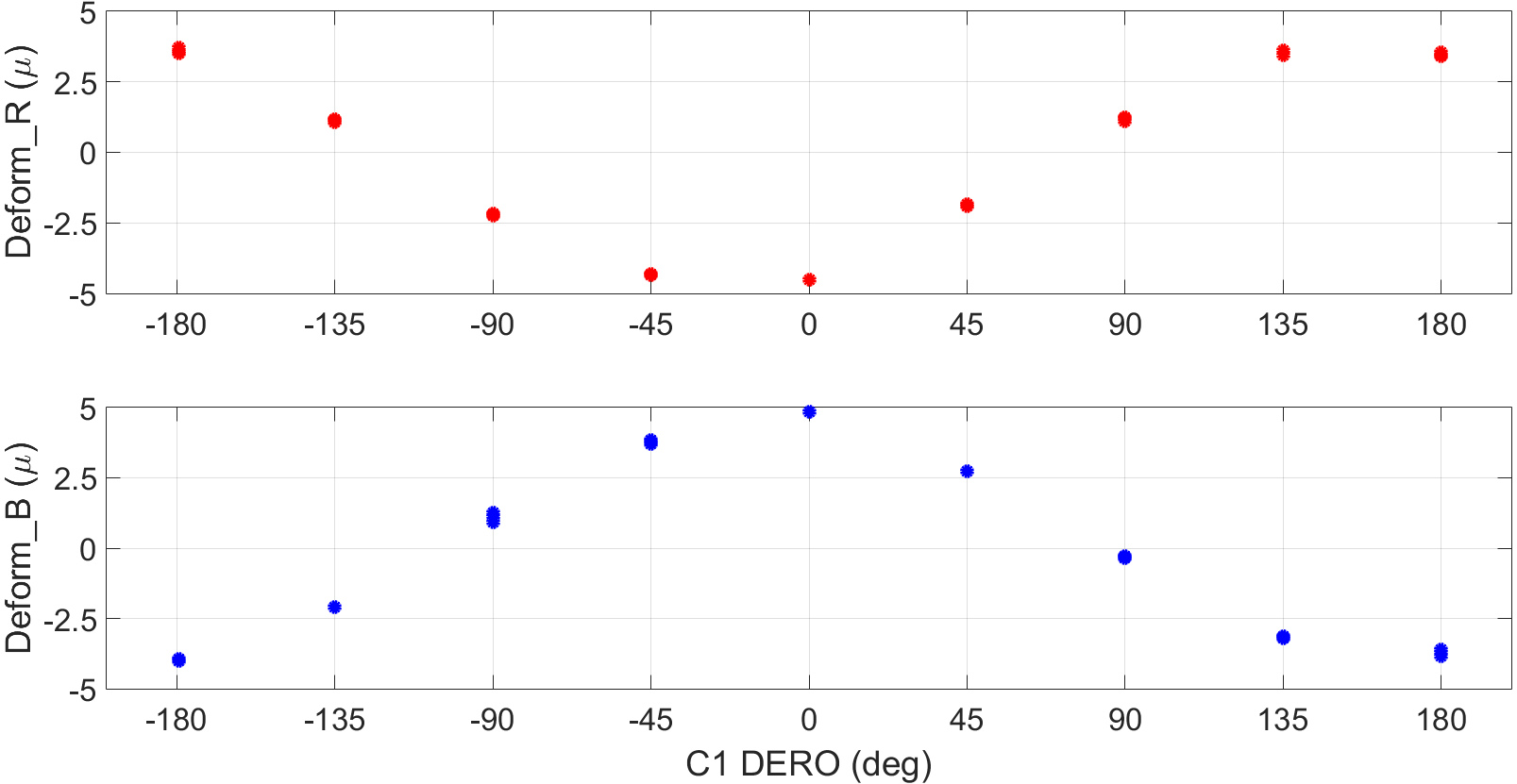}
\caption{The measured value of the TDS gravity sag depending on the tube elevation and the rotator angle (top panel; shift of spectral lines in Angstroms). Graph of the dependence of the spectral shift along the dispersion on the rotation recorded near the horizon ($10^{\degree}$ elevation) and expressed in microns (bottom panel)}
\label{fig:gravshift}
\end{figure}

The simulation of the spectrograph mechanics by the finite element method showed the stability of the mutual orientation of its optical components to within $5\mu m$ with an arbitrary change in the orientation of the instrument. Nevertheless, we performed a series of dedicated measurements to independently determine the character and magnitude of the gravity-induced deformations. At different telescope elevations, we measured the spectrum shift along the dispersion as a function of the Cassegrain rotator angle (Fig.\ref{fig:gravshift}). The elastic shift component has an amplitude of $\pm5\mu m$ (0.3\AA\ or 20~km/s for the $H_\alpha$ hydrogen line) and is the same in both channels. These shifts are mandatory for taking into account and must be eliminated from the determined Doppler shifts by the proper wavelength calibration of observations.

The hysteresis and non-reproducibility of gravitational deformations are far lower, less than $0.5\mu m$. The reproducibility of the slit setting was tested separately and found to be better than $0.7\mu m$. On the whole, we can conclude that the total uncorrectable mechanical error in the measured line shifts in the spectrum does not exceed $1\mu m$ (0.06\AA\ or 3~km/s at the $H_\alpha$ hydrogen line). The errors in measurements of the position of spectral lines associated with variations in the position of stellar images relative to the center of the spectrograph slit can be substantially higher than these errors of the mechanical origin.

In addition to the gravity sag, thermal deformations were also detected related possibly to some temperature-dependent stresses in the body of the spectrograph. Their value in the red channel is 0.05\AA/$1^{\degree}$C; in the blue channel the shift is remarkably smaller, $<$0.01\AA/$1^{\degree}$C.

Based on the presented data, we assess the reliability of measurements of Doppler line shifts during observations and develop the on-sky and day-time calibration procedures.

The system deformations also include the variable displacement of the TDS body as a whole due to the deflection of the support truss by means of which the instrument is attached to the telescope. With different orientations of the device the relative shift of the center of the slit to the pixel grid of the 4K-camera varies within a range of about 0.1--0.2~mm (1--2\arcsec). This complicates the acquisition of faint sources that are not visible in the viewer camera and, on the other hand, this flexure represents an additional source of displacement of sources on the slit during long exposures. Such effects have to be taken into account by the specific observational technique.

\section{Observing Protocol}
\label{sec:obs}
The TDS control software includes the \emph{andor-daemon} CCD camera setup and read-out software plus several Python scripts which control the mechanization of the spectrograph, launch of exposures, and also provide the operator's GUI. Two copies of \emph{andor-daemon} are running permanently because the interruption of communication with Andor cameras cancels the thermoelectric cooling. Startup commands and configuration or status parameters for detectors are transmitted over the socket connection. All information about the state of the telescope control system, the target and the observation circumstances is received through the EPICS bus\footnote{\url{https://epics.anl.gov/}}; this channel is also used to control the FSU and TDS mechanical motions.

The operator interface of the TDS is built around the spectrograph slit viewing camera GUI and is also used to perform corrections of the telescope pointing, positional angle of the slit and the focus of the telescope. It can also select the slit and put the wheel into the appropriate position, choose the blue channel dispersor, and start a series of spectral exposures (synchronously or separately by channels). In what follows we describe in detail the procedure for observations and calibrations of the instrument.

\begin{description}
\item[Centering the source.]
The 2.5-m telescope has a random pointing error of about 2\arcsec\ but for a number of reasons, uncompensated systematic deviations in the position of the object relative to the slit after pointing can reach 5--10\arcsec. Therefore, the target identification and centering (acquisition) is required before starting spectral exposures.

Stars as faint as $V\!\sim\!19^{\rm m}$ are visible in the slit viewing camera, fainter stars are centered on the slit in one of the following ways: (i) by centering on a brighter star also placed on the slit with the specific position angle, (ii) by applying pre-calculated offsets after centering the slit at some brighter star at a relatively small ($<\!1^{\degree}$) distance from the object, or (iii) by centering the target on the known pixel position in the 4K-camera of the Cassegrain port. The latter method has the advantage because some useful photometric information about the object of interest can be obtained, for example, the magnitude(s) in the SDSS system. The coordinates of the ``slit'' projection on the 4K-camera detector slightly vary depending on the target elevation and orientation of the Cassegrain rotator due to bending in the spectrograph support and therefore have to be calibrated against brighter stars each time after disassembly and re-installation of either the spectrograph or the wide-field camera on the telescope.

\item[Taking exposures of the source.]
Immediately after the target acquisition, the automatic guiding using the offset guider of the Cassegrain port of the telescope is started, which typically has an accuracy of $\sigma\!\sim\!0\farcs1$ during the entire measurement session. The guide star is selected in a ring with the radius of 10--15\arcmin\ around the scientific source. The viewer mirror of the spectrograph is flipped back, a desired slit is introduced and exposures are started. We usually take a series of 3--5 exposures each up to 20 minutes long; longer integration is disadvantageous because of the accumulation of cosmic ray hits.

Because of flexures in the suspension of the spectrograph during long series of observations (monitoring), the position of the object on the slit should be controlled between exposures. A slit viewing mirror is inserted into the path for adjusting relatively bright sources or, otherwise, an FSU mirror is deployed and short exposures are taken with the 4K-camera. It takes 16~seconds to insert or remove the FSU/TDS mirror; the TDS slit viewing mirror and the slit wheel moves to a desired position within 1--2~sec.

The \emph{expose} script performs exposures, saves raw spectrum files in the FITS format, and stores in the database with the observing log all the relevant parameters of the target and observations. These include the detector settings and the moment of the start of an exposure, identifiers of the target and its research program, coordinates of the source from a catalog, and current readings of the mount axes (parallactic and the positional angles of the Cassegrain rotator, as well as target elevation are recorded at the moments of the beginning and the end of the exposure), the positions of the TDS and FSU drives, meteorological parameters and temperatures of the telescope optics and spectrograph elements. It takes about 25~sec to readout the frame and save all this information.

\item[Wavelength calibration.]
It is sufficient to obtain arc spectra with a high signal-to-noise ratio once a month or after interference in the optical system. Corrections to the wavelengths associated with instrument deformations for faint objects can be calculated from the position of emission lines in the night sky background. For brighter objects, the exposures are insufficient to well expose sky background lines. Therefore, if one needs precise radial velocities for a given target, it is recommended to obtain a short (100~s exposure) arc spectrum before and after obtaining the spectra of the science target.

\item[Spectrophotometric standards] are measured before or after science targets at a similar airmass. We use the stars from the list of standards of the European Southern Observatory\footnote{\url{https://www.eso.org/sci/observing/tools/standard/spectra/stanlis.html}}, among which there are white dwarfs with smooth and well-measured spectra. To perform the correction for telluric absorption due to the Earth's atmosphere, any star of the A0V spectral type can be used which possesses no spectral/chemical peculiarities and has an interstellar absorption $A_V<0.2^{\rm m}$.

\item[Flat field calibration] is performed using a continuum spectrum LED source once a day. On clear sky, the twilight background spectra are recorded approximately once a week to clarify the shape of the slit function (see next section).

\item[Dark frames] are accumulated approximately once a season, usually on cloudy nights. Since the growth of counts in hot pixels is nonlinear and reaches saturation, a series of exposures on a grid of the most frequently used exposure times $T_{\rm exp}=0,$ 300, 600 and 1200~sec is made, 10 frames per each exposure time to eliminate cosmic ray hits and improve statistics.

\end{description}


The described procedure is currently performed by the instrument operator/observer; in the future we plan to automate the object acquisition and calibration measurements.

\section{Spectral Data Processing}
We show the real measured characteristics of the spectrograph in the bottom of Table~\ref{tab:param}. In practice, the proper focusing of the spectrograph yields a spectral resolution only 10--20\%\ worse than theoretical value determined from the geometric width of the image of the  1\arcsec\ slit. Similar values of spectral resolution were obtained by fitting the twilight spectra against a template obtained by a convolution of the reference spectrum of the Sun\footnote {\url{https://noirlab.edu/public/images/noao-sun/}} with a model of a variable width instrument \emph{line spread function} parameterized by the 4-th order Gauss-Hermite function. The images are practically symmetrical over the entire wavelength range, which is proven by nearly zero values of the $h3$ Hermite coefficients.

To reduce raw spectra, we developed a {\sc python}-based data processing package, which can by run in automatic and interactive modes and relies on the metadata in the headers of the input FITS files. The processing includes the following components:
\begin{description}
\item[Combining dark frames] by the median averaging.

\item[Dark frame subtraction] is performed after a linear interpolation of the combined dark frames to the exposure time of a given science or calibration frame.

\item[Removal of cosmic ray hits] is performed when needed using the LAcosmic \citep{2001PASP..113.1420V} package.

\item[Wavelength calibration] of the slit image is performed using the spectrum of the hallow cathode lamp. For calibration, we compiled a list of the non-blended arc lines distributed across the spectral range. The positions of the lines in each CCD line are determined by fitting with a Gaussian profile, after which the position for each line is fitted with a polynomial that simulates the observed curvature of the slit image. For convenience, dispersion curves are computed for each CCD row using these averaged positions and fitted with a 5th degree polynomial for the $B$ and $G$ channels and a 7th degree polynomial for the $R$ channel. 

Typical residual scatter is 0.20\AA\ in $B$, 0.07\AA\ in $G$, and 0.01\AA\ in $R$. The lower calibration accuracy in the blue channel is because of the weaker spectral feature, fewer number of non-blended lines, and a lower spectral resolution.

\item[Rebinning the data] onto a uniform wavelength grid is performed by linear interpolation of frames according to an individual dispersion curve for each line of the detector, which simultaneously corrects the curvature of the slit image.

\item[Sensitivity variations along the slit] are corrected by calculating the ``flat field'' from the spectrum of the twilight sky and from the spectrum of the continuum LED source. Flat fields with the LED lamp are recorded faster and more uniformly in noise, since they do not have strong absorption lines that are present in the sky spectrum. The uneven illumination of the slit by the LED source is seen as the illumination gradient $\sim\!10$\% along the slit. To eliminate this effect, a low-frequency correction is calculated using flats obtained from a twilight sky. Flat fields not only correct dust shadows and slit imperfection, but also effectively eliminate fringes observed in the red channel.

\item[Extraction of spectra] is performed in a GUI using two different algorithms:
\begin{enumerate}
\item 
A standard (linear) extraction by co-adding the counts within a given aperture with the subtraction of the background, which is determined by the areas further away from the source.
\item An optimal extraction according to the algorithm described by \cite{1986PASP...98..609H}, which consists of calculating the spatial profile of the spectrum and then co-adding the counts weighted with this profile.
\end{enumerate}


The use of optimal extraction makes it possible to remove the remaining traces of cosmic rays and is advisable for relatively faint objects, for which it is possible to obtain the signal-to-noise ratios 20-30\% higher than with a linear extraction. For brighter objects, standard and optimal extractions lead to almost identical results.

\item[Wavelength correction] Simultaneously with the spectrum extraction, the  wavelength correction can be calculated and taken into account using night sky lines. In the blue channel the line 5577\AA\ and the lines of mercury Hg\,I 4046.57, 4358.34, 5460.75\AA\ from the light pollution are used. In the $R$ channel, many oxygen and hydroxil lines, and also the sodium Na$D$ doublet are available for calculating the correction.

\item[Calculation of the spectral response] is carried out on using spectra of standard stars with known energy distributions, whose spectra are obtained close in time to the science target and at the same airmass. Correction for the system flux response is performed according to the calculated ratio of the extracted spectra of the standard stars to the published spectral energy distributions. The obtained ratio reflects the spectral efficiency of the instrument for the entire measurement path from the upper boundary of the atmosphere to the received photoelectrons of the signal of the detectors. 

\item[Flux calibration] is performed by dividing the instrumental flux, measured in counts per second, by the response curve. The final values have the dimension [erg/(cm$^2\!\times$s$\times $\AA)].

\end{description}

All data processing history, parameters of algorithms and control information, as well as barycentric corrections for wavelengths, are stored in FITS headers of the output files containing the calibrated spectrum. Spectrum extraction parameters and results are written into additional FITS extensions. The merging of the results of blue and red channels into a single spectrum is performed, if necessary, separately from the standard processing described above.

In addition to the {\sc python} data processing package, an alternative version of the TDS data processing pipeline is available for Interactive Data Language (IDL, commercial software package) or its free analogue, GNU Data Language (GDL). This package is free software licensed under the GPLv3\footnote{\url{https://bitbucket.org/chil\_sai/mosifu-pipeline}} and is based on the algorithms from the automatic data processing pipelines of the MMIRS and Binospec spectrographs \citep{2015PASP..127..406C, 2019PASP..131g5005K} installed at the 6.5-meter MMT telescope in Arizona. The pipeline is a universal long-slit spectra processing system, where we added a block with a configuration for the TDS. In general, the algorithms used in it are similar to those described above, but there is a number of differences: (a) to construct a wavelength solution, we fit the positions of arc lines over the entire frame by a two-dimensional polynomial; (b) an alternative algorithm for cleaning cosmic ray hits is implemented that uses statistical analysis of several spectral frames simultaneously (if available); (c) any star of the spectral type A0V can be used as a spectrophotometric standard provided that it does not have substantial foreground dust extinction on the line of sight (up to $E(B-V)=0.05^{\rm m}$) and strong chemical composition peculiarities -- this provides additional flexibility while choosing a star near the object, since such stars are quite frequent in the sky; (d) the procedure for subtracting the night sky is optimized for faint objects and is applied before linearization and correction of the geometry of a two-dimensional spectrum \citep{2011ASPC..442..143K}; (e) the final spectrum is optionally corrected for telluric absorptions.

\section{Throughput of the System}


Observations through a wide slit allow the spectrophotometry to be carried out at the cost of reduced the spectral resolution. These observations, carried out quasi-simultaneously at different air masses, make it possible to separately reconstruct the current atmospheric transmission curve and the response function of the instrument. To achieve high resolution and photometric accuracy, two measurements are required, with a wide and a narrow slit.


Bold line in Fig.~\ref{fig:dichro} shows an example of recovering the throughput curve of the spectrograph with the atmospheric extinction taken into account as described above.
We used the standard stars BD+75d325 and BD+28d4211 observed with a 10$^{\prime \prime}$ slit at airmasses $M=2.0$ and $M=1.03,$ respectively. The presence of narrow features in the throughput curve located in the region between $ H_\gamma$ and $ H_\delta$ requires high quality and resolution of the spectra used for calibration purposes.
The throughput uncorrected for the atmospheric extinction measured for the star at the altitude of $75^{\degree}$ reaches 35\% in the red, 22\% in the green and 20\% in the blue channel. Similar values were obtained with other standard stars. 


The reduced sensitivity of the blue channel as compared to the red one prompted to independently calculate the expected system transmission near the central wavelengths of the channels for the object near the zenith. For the extinction of the atmosphere, the median value at the CMO were taken as 0.13 in $R$ and 0.25 in $B$ \citep{2016MNRAS.462.4464K}. For the telescope mirrors M1 and M2, we took the reflection coefficients of the mirrors averaged over the surface reduced by 8\% and 5\% to take into account the aging of coatings \footnote{The KGO is monitoring the reflectivity of the mirrors of the 2.5-m telescope on a relative scale.}. We took the values of the throughput, reflectivity, and quantum efficiency provided by the corresponding manufacturers for the remaining elements.


The final value of the theoretical transmission of the entire path near the zenith in the $R$ channel is 37\%, close to the measured value. At the same time, the transmission measured in the $B$ mode (blue channel) is only 73\% of the expected value of 27.6\%. The transmission in the $G$ mode (blue channel) at the wavelength of the maximum efficiency of the grating (505~nm) turned out to be 10\% higher than in $B$, although it should not exceed it according to the calculations. The reasons for these inconsistencies remain to be determined.


Taking into account the estimated transmission of the feeding optics of the spectrograph (the telescope and the folding diagonal mirror), the quantum efficiency of the TDS itself including the CCD detectors reaches 65\% in the $R$ channel, 47\% in the $B$ channel, and 50\% in the channel $G$ near the central wavelengths of the dispersors. These characteristics of the spectrograph are provided in Table ~\ref{tab:param}.

\begin{figure}
\includegraphics[width=\hsize]{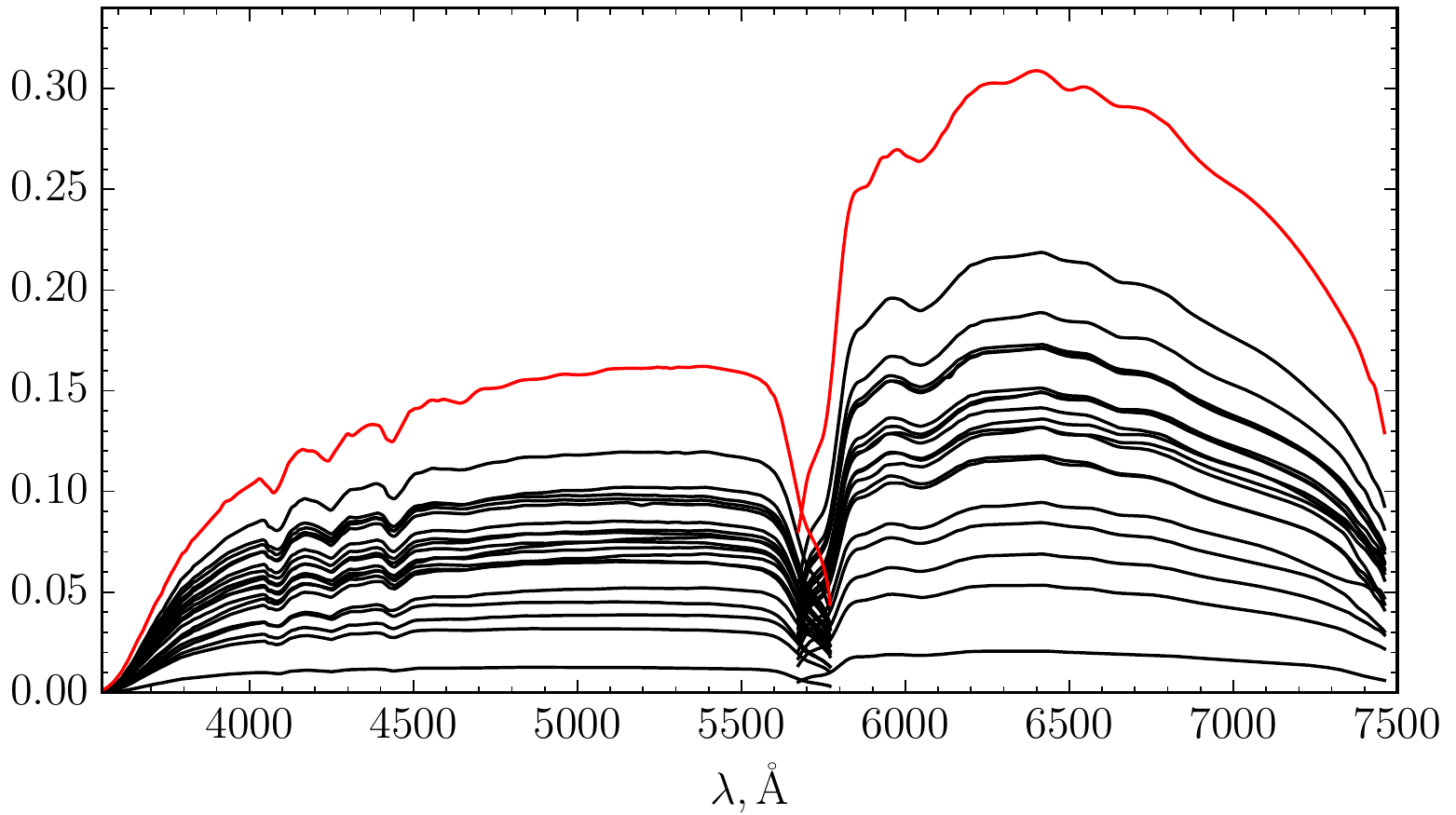}
\caption{A set of throughput curves without correction for the atmospheric extinction in the $B$ and $R$ channels measured with the spectrophotometric standard star BD+75d325. The upper (red) curves are measurements with the 10\arcsec-wide slit at an altitude of 30${\degree}$, while the rest corresponds to the 1\arcsec-wide slit at the altitudes between 30 and 50$ {\degree}$.
}
\label{fig:slitloss}
\end{figure}

When observing standard stars with a narrow slit, only small-scale transmission inhomogeneities of the throughput curve can be accounted for, while the total transmission remains undetermined due to slit losses, which can vary because of the wavelength-dependent point source profile on the slit and atmospheric dispersion. The latter effect can be eliminated by setting the slit along the parallactic angle as noticed above, if a given observing program permits. However, the total loss in the atmosphere and on the slit can vary over a very wide range (see Fig.~\ref{fig:slitloss}).


To estimate the expected signal at the maxima of the sensitivity of each channel, we calculated the magnitudes of A0V class stars, which, without the atmospheric extinction, yield a signal of 1~$e- s^{-1}$~pixel$^{-1}$ for the wavelength sampling of the extracted spectrum: $mag_1(B)=17.6^{\rm m} $, $mag_1(G)=16.8^{\rm m}$, $mag_1(R)=17.2^{\rm m}$. When estimating the exposure time for science targets, these values should be corrected for the expected values of extinction, airmass, and an estimate of the spectrograph slit loss due to the atmospheric seeing quality.

\section{First Observations and Results}

Due to the good throughput of the system resulting from the simple and efficient optical design, and the average spectral resolution 1.5--2 times that in existing large spectroscopic surveys (SDSS, LAMOST), the TDS occupies a specific niche which it shares with typical low-resolution spectrographs at telescopes with a mirror size of the order of 3--3.5~m (Shane Telescope of the Lick Observatory, NTT of the European Southern Observatory). In a reasonable exposure time (2--4 hours), TDS allows one to study in detail astrophysical objects 2--3 magnitudes fainter than those included in the samples of modern spectroscopic surveys (i.e. quasi-point objects up to $21.5^{\rm m}$), or to obtain spectra with a high signal-to-noise ratio for slightly brighter objects for which only poor-quality spectra are available in surveys.

Point-like and extended sources of various types are studied with the spectrograph: variable stars, nuclei of planetary nebulae, active nuclei and disks of galaxies, quasars, supernovae etc. Regular observations are carried out according to the programs for monitoring the microquasar SS433 (see below), novae \citep {2020ATel14004....1S}, spectroscopy of objects from the Spektr-RG/e-Rosita \citep{Dodin2020_SRG} mission, active galactic nuclei \citep{ilic2020}, symbiotic stars (see below), white dwarfs \citep{Pshirkov2020} and planetary nebulae \citep{arkhipova2020n-3-75327450852}. We also follow-up galaxies with intermediate mass black holes \citep{2018ApJ...863....1C} and study the environment of rare giant galaxies of low surface brightness \citep{2018MNRAS.481.3534S, 2019MNRAS.489.1463O} which happen to be accompanied by even rarer type ``compact elliptical galaxies'' satellites \citep{2009Sci...326.1379C, 2015Sci...348..418C}.

Some results illustrating the capabilities of the TDS are presented below. 

\subsection{Monitoring of SS433}

\begin{figure*}
\includegraphics[width=\textwidth]{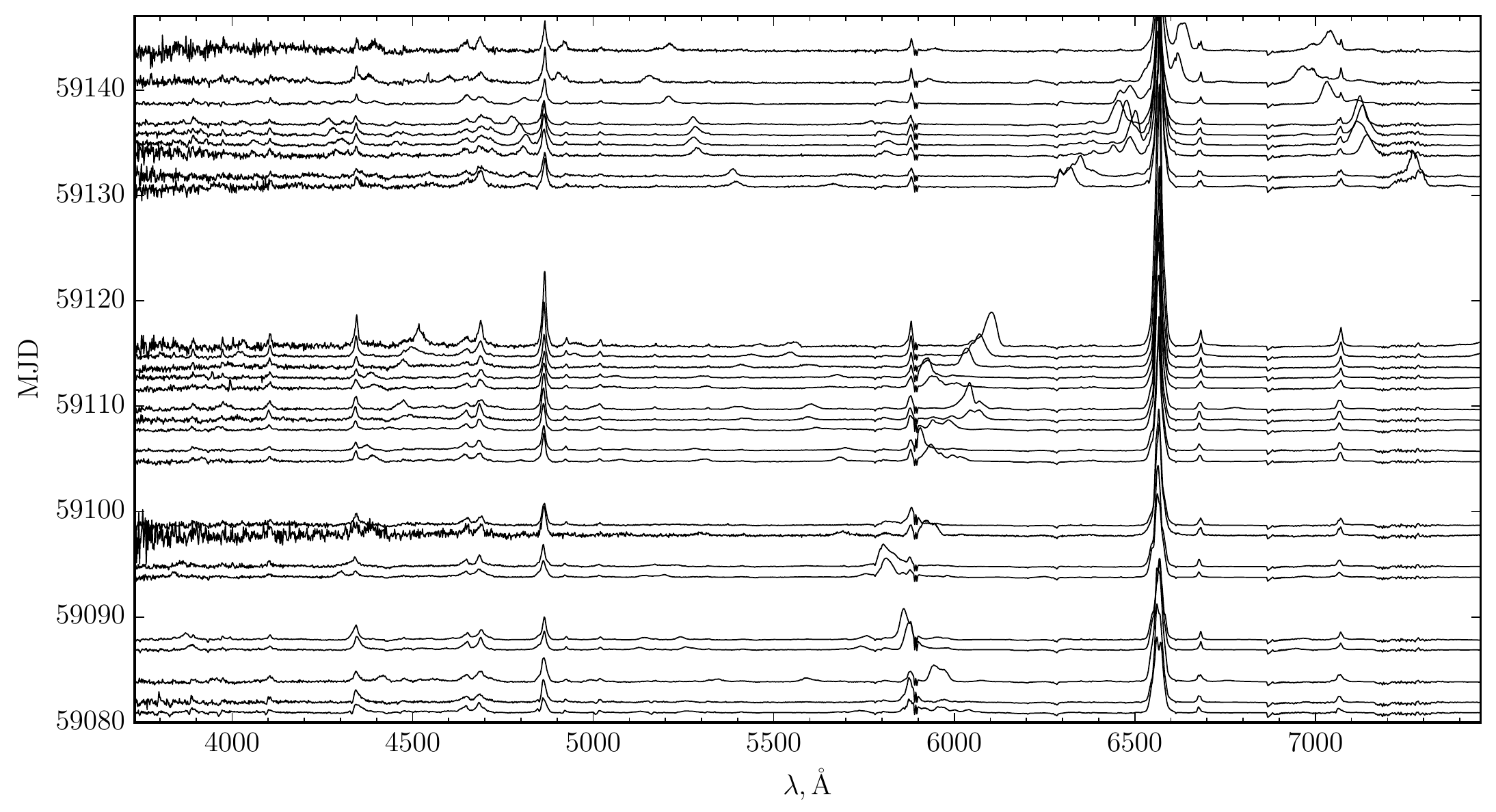}
\caption{Evolution of the spectrum of SS433 between August 19 and October 21, 2020. The exposure time varies from   300 to 900~s, the atmospheric transparency and seeing are variable.}
\label{fig:ss433}
\end{figure*}

TDS has been used to perform several monitoring sessions of the Galactic microquasar SS433 to study the variability of line profiles and radial velocities in the spectrum. SS433 has been studied in detail at SAI MSU for a long time \cite{2018ARep...62..747C,2020NewAR..8901542C}. Fig.~\ref{fig:ss433} presents time evolution of spectra obtained by TDS in August--October (Modified Julian Dates of exposures are shown in ordinate axis), 2020. In this Figure, the spectra from two channels are merged, the continuum is subtracted, intensities are in arbitrary units. The brightness of the object allows to perform monitoring almost every night under any weather conditions. These measurements are important to trace possible evolutionary changes of the parameters of this massive X-ray binary system which is observed at a special phase of the secondary mass transfer. For some reasons, a common envelope has not been formed and the loss of mass and angular momentum from the system occurs from a precessing super-critical accretion disk around a relativistic object (most likely, a black hole). Thus, observations of SS433 at a critical evolutionary stage represent an undoubtedly important task.

\subsection{Studies of Symbiotic Stars}
\begin{figure*}
\includegraphics[width=\textwidth]{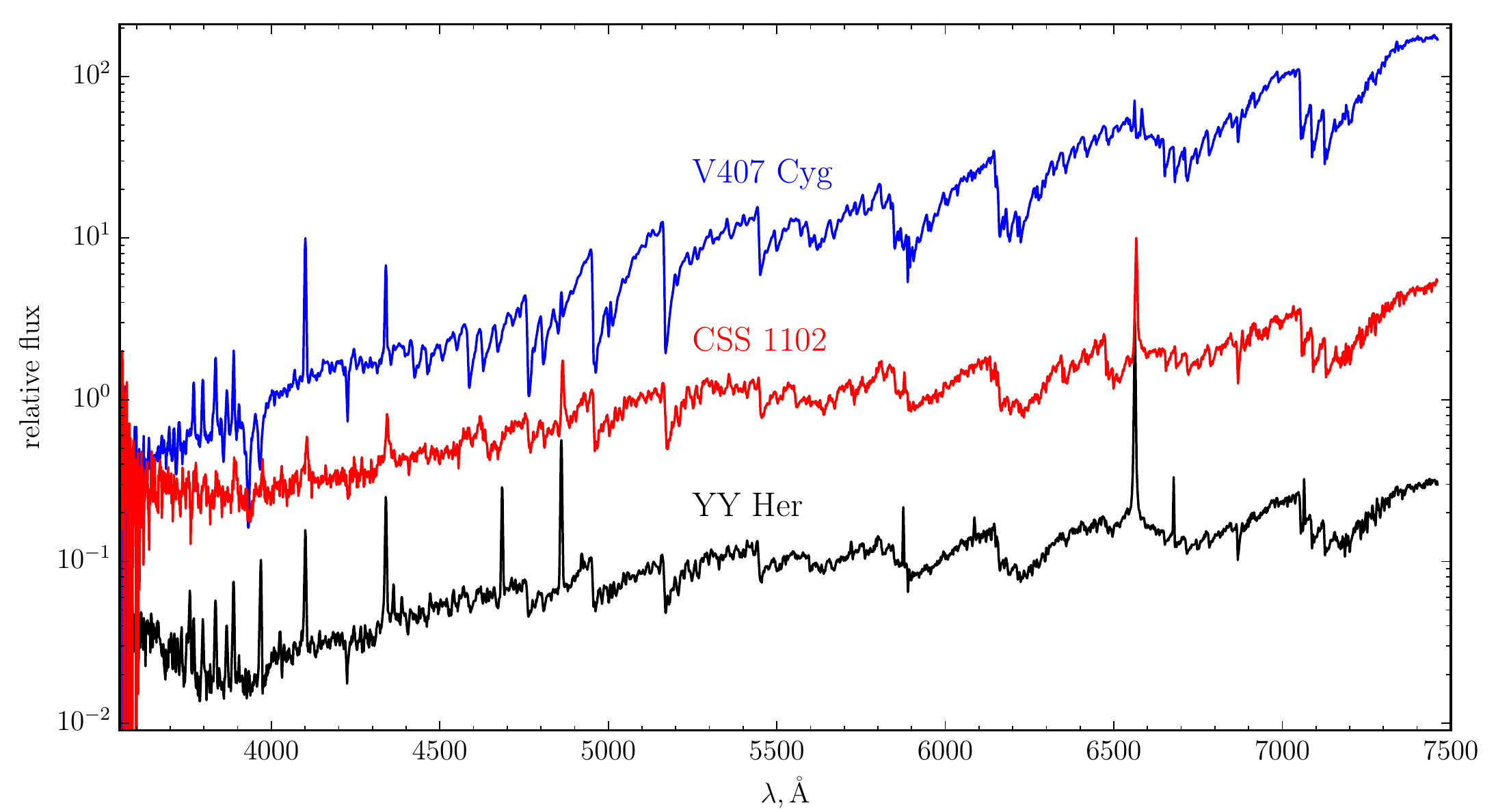} 
\caption{The CSS~1102 star spectrum (MJD$=59092.75,$ $T_{\rm exp}\!=\!100$ sec, $V\!\sim\!13^{\rm m}$) in comparison with the V407~Cyg ($T_{\rm exp}=500$ sec, $V\!\sim\!15^{\rm m}$) and YY~Her ($T_{\rm exp}=100$ sec, $V\!\sim\!11^{\rm m}$) stars.}
\label{fig:example1}
\end{figure*}

In Fig.~\ref{fig:example1} we present the spectra of a few symbiotic stars V407~Cyg, CSS~1102 and YY~Her obtained at different stages of activity. The symbiotic star V407~Cyg is currently in a passive stage and exhibits a typical spectrum of an oxygen-rich Mira-type star, with a Balmer decrement characteristic of this class of objects (H$_{\delta}$ is the strongest emission line in the spectrum). The spectrum of the classical symbiotic star YY~Her was obtained in a quiescent state. In addition to bright emission lines of H\,I, He\,I, He\,II and others, a nebular continuum is clearly visible that weakens the molecular bands of the cold component in the blue region. The TDS allows one to obtain spectra in the region of the Balmer break which is important for studying both the symbiotic nebula and the accretion disk which can appear at certain stages of outbursts in some systems.

The middle spectrum in Fig.~\ref{fig:example1} shows a relatively poorly studied star CSS~1102. The object was discovered during the survey by \citet{CSSobzor} as an S3 star. In August 2020, a message appeared about the possible symbiotic nature of this source\footnote{\url{https://www.aavso.org/aavso-alert-notice-719}}. Our spectral observations showed that in the spectrum of the star, in addition to the TiO and ZrO molecular bands, there are bright emission lines of hydrogen, and the Balmer decrement has a value uncharacteristic for miras (see the spectrum of V407~Cyg in the same figure above). The spectrum of the object also shows the lines of neutral helium and traces of forbidden lines [Ne\,III]. This undoubtedly allows the object to be classified as a symbiotic star. Our photometric observations carried out on August 26, 2020 with the 60-cm telescope of CMO detected a rapid brightness variability (flickering) in the CSS~1102 system with a characteristic time of about 30~min and an amplitude of $0.1^{\rm m}$ in the~$B$ filter (Tatarnikova et al., in preparation). This is a rare type of variability in symbiotic stars, probably associated with the presence of accretion disks and jets in these systems. The vast majority of such objects belongs to a rare subclass of recurrent symbiotic novae. 

\subsection{Transients and Flaring Stars}



\begin{figure}
\includegraphics[width=0.47\textwidth]{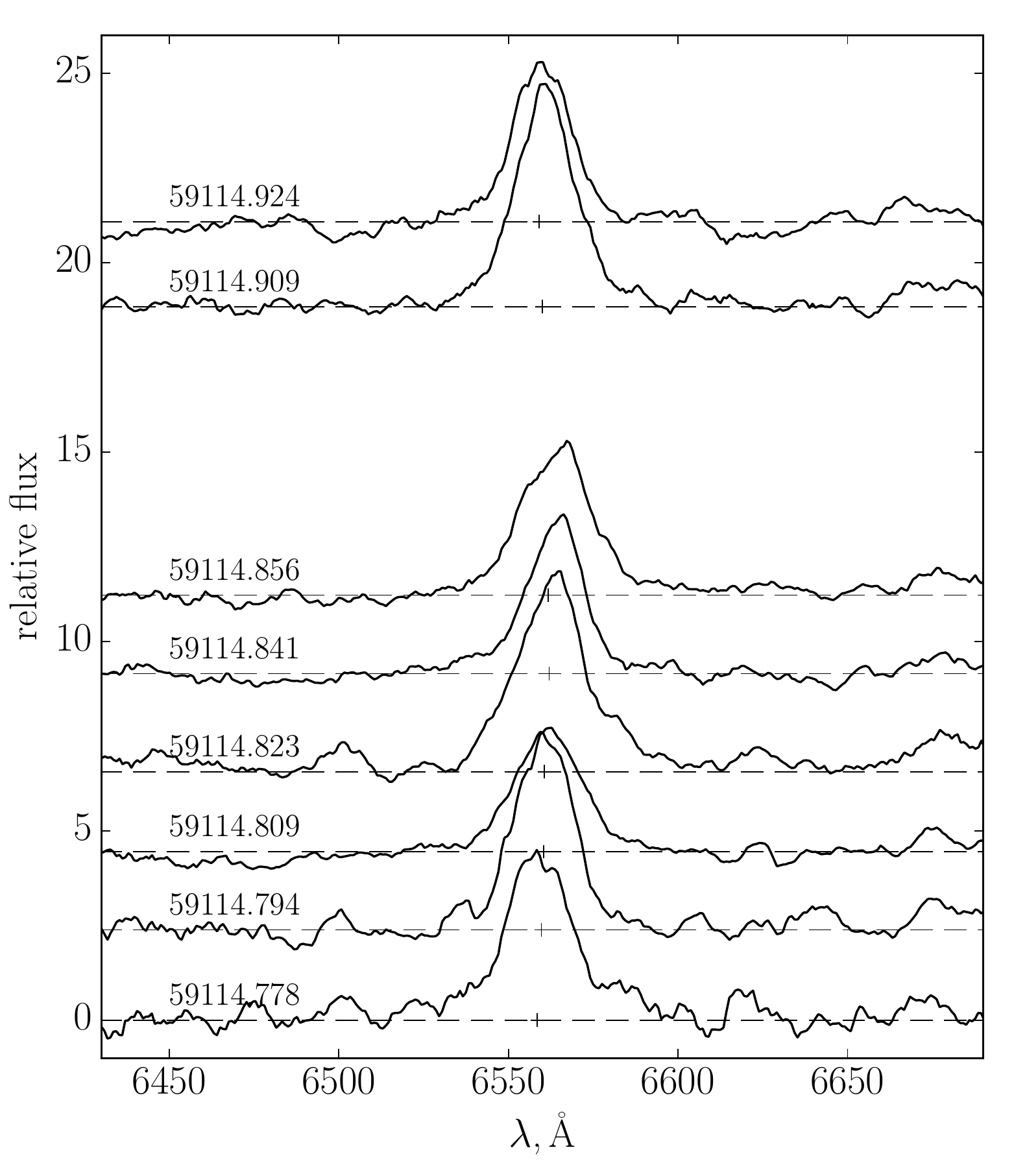} 
\caption{Variability of the H$\alpha$ profile in the spectrum of the dwarf nova SAI-V J004051.59+591429.7 at the minimum of brightness ($R_c\!\sim\!20^{\rm m}$). The centroid of the line is marked with a cross. The exposure time was 1200~sec, the signal-to-noise ratio in continuum is $\approx2.5$ per pixel. The spectra are smoothed by moving averages over 10 points. On the left, the heliocentric MJD moments in the middle of the exposure are shown.}
\label{fig:example2}
\end{figure}

The dwarf nova SAI-V J004051.59+591429.7 was discovered\footnote{\url{http://www.sai.msu.ru/new_vars/}} at the CMO SAI MSU during photometric monitoring with the 60-cm telescope of another variable star. In 2019, two outbreaks of this dwarf  nova were observed. Both flares correspond in characteristics to normal dwarf nova flares. The spectra of the variable ($R_c\!\sim\!20^{\rm m}$) obtained with the 2.5-m CMO telescope using the TDS are characteristic of dwarf novae at the minimum brightness. The emission lines of the Balmer series of hydrogen are clearly visible. The radial velocities determined from the $H_\alpha$ emission show regular variability associated with the orbital motion in the binary system (see Fig.~\ref{fig:example2}). The orbital period is more than 3.5 hours; further spectral monitoring will make it possible to refine it.


\begin{figure*}
\includegraphics[width=\hsize]{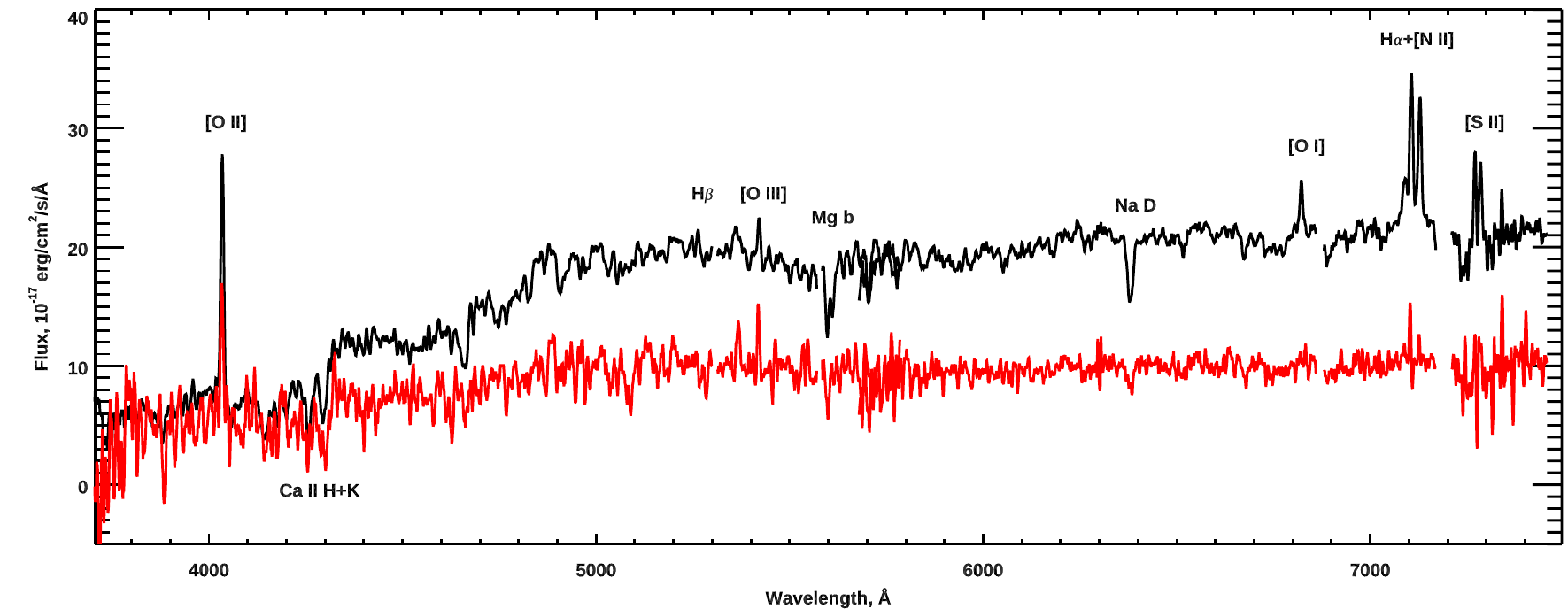}
\caption{Spectra of the central part of the low surface brightness galaxy Malin~1 obtained with a total exposure of four hours, $g = 18.72^{\rm m}$ (the black line), and of one of its compact elliptical satellites, $g = 19.94^{\rm m}$ (the red line). The spectral fluxes of the satellite are increased by a factor of three for clarity. The main absorption and emission lines are marked. Both objects host active galactic nuclei.}
\label{fig:example3}
\end{figure*}

\subsection{Absorption- and Emission-Line Spectra of Galaxies}


The TDS makes it possible to study both emission- and absorption-line spectra of galaxies and active galactic nuclei. The intermediate spectral resolution $G$ mode with the grism in the blue channel covers nearly all the most frequently used absorption lines in the optical spectra of galaxies, which are used to measure stellar kinematics and assess the properties of the stellar populations (age, chemical composition). At the same time, the available spectral resolution makes it possible to measure the velocity dispersions of stars up to 35--40~km/s in galaxies with old stellar populations with a signal-to-noise ratio of about 15 per pixel (or up to 70--80~km/s when using the low resolution mode, see \citet{2020PASP..132f4503C}, where the theoretical calculation of uncertainties of the parameters of stellar kinematics from absorption spectra is discussed, which we used for our estimates).


Thus, it becomes possible to study the internal dynamics of relatively low-mass dwarf galaxies that are inaccessible for study with other Russian telescopes, including BTA because of to insufficient spectral resolution. At the same time, the red channel provides a sufficient resolution of the H$_\alpha$ emission line profile, which makes the TDS an excellent tool for studying active galactic nuclei \citep[see e.g.][]{ilic2020}, including those containing intermediate mass black holes \citep{2018ApJ...863....1C} where the broad line component is rather weak and relatively narrow (150--250~km/s). The blue channel for such objects will be sufficient for measuring relatively low stellar velocity dispersions in their host galaxies (40--70~km/s).


Fig.~\ref{fig:example3} shows the spectra of a giant galaxy of low surface brightness Malin~1 (black line) and its compact elliptical companion (red line). Both galaxies host faint active galactic nuclei. Despite the satellite's magnitude of 20~mag in the SDSS $g$ filter, the spectrum clearly shows not only the emissions associated with the active nucleus, but also the absorptions from its stellar population.

\section{Conclusion}
\label{sec:conclusion}

Results of pilot observational programs with the TDS have justified the demand of the high throughput of the instrument along with a spectral resolution sufficient to measure both integral intensities and spectral line profiles for non-stationary and extragalactic sources. Among long-slit spectrographs on intermediate-sized telescopes briefly mentioned in the Introduction, no TDS analogues (a fully dioptric system with dichroic beam splitter and holographic dispersors) were found. Therefore, direct comparison of characteristics of TDS with other instruments is difficult. 

Commissioning and science verification observations with the spectrograph suggest that the efficiency of the instrument red channel generally corresponds to the theoretical value calculated using the transmission curves of the atmosphere, and throughput of the optics of the telescope and spectrograph, the efficiencies of the dispersor and the detector. By this parameter, expectedly, TDS stays among high-throughput world-class spectrographs. However, there are still several issues to be studied in detail in the future. In particular, we need to understand the lowered blue-channel efficiency and the rapid decline of the instrumental respond towards 350~nm that hampers spectrophotometric measurements of objects near the Balmer break. 

The spectrograph's efficiency can be further improved. For example, it would be worthwhile producing the beam splitter with complex dielectric covering by a manufacturer specialized in astronomical instruments. This could help to suppress the effect of ``waves'' in the reflection and throughput characteristics and to improve the reconstruction of the throughput curve, which is especially desirable in the region of high-order hydrogen lines in the blue spectral channel of the spectrograph.

A significant decrease in the scattered light and additional 6\% in the efficiency could be secured by changing the standard detector windows by AR-coated ones in the working spectral range of the device.  
A multi-layer dielectric coating of the feeding diagonal mirror could produce a similar gain in efficiency. Finally, the improvement of the telescope tracking (due to its pointing model imperfections) could reduce losses on the slit which can be significant for faint sources observed with long exposures.

As an instrument for spectrophotometric research, TDS has a great potential. The first results have demonstrated that throughput curves reconstructed using different standard stars are consistent within 2\%\ or better. However, secure spectrophotometric observations of various sources require a separate study of the stability of estimates obtained from observing standard stars over multiple nights with photometric conditions.

A limited number of spectra to measure radial velocities have been obtained with TDS so far. The study of Pshirkov et al. (\citeyear{Pshirkov2020}) suggested that using dedicated methods radial velocities can be reliably detected within the range 10--20~km/s even for faint objects with relatively featureless spectra. However, more observations are required to confirm this estimate. 

The presented characteristics of the TDS spectrograph of the 2.5-m telescope of CMO SAI MSU and the results of the first year of its operation demonstrate that it is an optimal instrument for prompt classification and long-term monitoring of non-stationary and transient astronomical objects. The use of TDS enables us to collect spectra of faint objects down to  $\sim\!20^{\rm m}$ in 1--2 hours of integration with a signal-to-noise ratio of five or more per one pixel of the detector and a resolution power of $R\!\sim\!1500$--3000 depending on the spectral setup.

\acknowledgements

The construction of the spectrograph was supported by RSF grants 16-12-1059 (design and manufacturing of the red channel optics) and 17-12-01241 (optics and camera of the blue channel). The work of SAP, NISh, AAB, SGZh, AMT, AVD, KAP, IVCh was partially supported by the Leading Scientific School of MSU ``Physics of stars, relativistic objects and galaxies''. SGZh, AVD and KAP are also supported by the Ministry of science and higher education of Russian Federation under the contract 075-15-2020-778 in the framework of the Large scientific projects program within the national project ``Science''  (observations, data reduction and analysis). Research made with the 2.5-m telescope of CMO SAI MSU has been supported by the Program of Development of Lomonosov Moscow State University. The authors acknowledge Drs. V.L.~Afanasiev and A.V.~Moiseev from SAO RAS for valuable consultations and advices during the construction of the instrument adn A.A.~Tokovinin for valuable remarks.

\label{lastpage}

\bibliographystyle{mypazh}
\bibliography{tds4sai}
\end{document}